\documentclass{aa}

\usepackage{graphicx,amssymb}
\usepackage{txfonts}

\def\gr{\hbox{ \raisebox{-1.0mm}{$\stackrel{>}{\sim}$} }}
\def\kr{\hbox{ \raisebox{-1.0mm}{$\stackrel{<}{\sim}$} }}

\begin{document}


\title{Prospects for multiwavelength polarization observations of 
GRB afterglows and the case GRB 030329}

\author{
S. Klose\inst{1}, 
E. Palazzi\inst{2}, 
N. Masetti\inst{2},
B. Stecklum\inst{1},
J. Greiner\inst{3},
D. H. Hartmann\inst{4},
H. M. Schmid\inst{5}
}

\offprints{S. Klose (\textit{klose@tls-tautenburg.de})}

\institute
{ Th\"uringer Landessternwarte Tautenburg, 07778 Tautenburg, Germany
  \and 
  Istituto di Astrofisica Spaziale e Fisica Cosmica, Sezione di Bologna, CNR,
  Via Gobetti 101, I-40129 Bologna, Italy
  \and
  Max-Planck-Institut f\"ur extraterrestische Physik, 85741 Garching, Germany
  \and
  Clemson University, Department of Physics and Astronomy, Clemson,
  SC 29634, USA
  \and
  Institut f\"ur Astronomie, ETH Z\"urich, 8092 Z\"urich, Switzerland
}

\date{\today }

\authorrunning{Klose et al.}
\titlerunning{Polarization of GRB afterglows}

\abstract{We explore the prospects for simultaneous, broad-band,
multiwavelength polarimetric observations of GRB afterglows. We focus on the
role of cosmic dust in GRB host galaxies on the observed percentage
polarization of afterglows in the optical/near-infrared bands as a function of
redshift.  Our driving point is the afterglow of GRB 030329, for which we
obtained polarimetric data in the $R$ band and $K$ band simultaneously
$\sim$1.5 days after the burst. We argue that polarimetric observations can
be very sensitive to dust in a GRB host, because dust can  render the
polarization of an afterglow wavelength-dependent. We discuss  the
consequences for the interpretation of observational data and  emphasize the
important role of very early  polarimetric follow-up  observations in all
bands, when afterglows are still bright, to study the physical properties of
dust and magnetic fields in high-$z$ galaxies.

\keywords{gamma rays: bursts -- polarization; dust -- extinction}
}

\maketitle

\section{Introduction}

Various independent studies of GRB afterglows have shown that in the optical
bands  the degree of linear polarization  is usually rather low, in the order
of 1 to 2\% (for references see, e.g., Bj\"ornsson 2003; Covino et al. 2002,
2003; Greiner et al. 2003a; Lazzati et al. 2003; Masetti et al. 2003). If this
turns out to be typical of optical afterglows (of long bursts), then the
contribution of the dust in GRB host galaxies  to the observed polarization
properties of the afterglows may not be negligible. While dust in a GRB host
galaxy will also manifest itself by changing the spectral energy distribution
of the afterglow light away from an original power-law (e.g.,  Palazzi et
al. 1998; Ramaprakash et al. 1998),  polarization provides another independent
method to constrain the existence of a dust component. If the sources
responsible for the long bursts are located in star-forming regions, cosmic
dust in the GRB environment will imprint a signal on the polarization
properties of the afterglow light in the optical/near-infrared (NIR)
bands. One might then imagine a situation in which the polarization by
intervening dust (which might be time-independent) can dominate over the
intrinsic polarization of the afterglow light. Motivated by our simultaneous
optical and NIR polarimetric observations of the afterglow of GRB 030329, we
explore here how dust in the GRB hosts may affect the polarimetry of the
afterglow.  Thereby, we concentrate on the wavelength dependence of the degree
of linear polarization $P$. While the intrinsic polarization of the afterglow
light  is expected to be due to synchrotron radiation and, thus, wavelength
independent, any additional polarization by dust in the GRB host can introduce
a wavelength dependence to the measured $P$. Having this in mind, we basically
focus on the redshift effect, which modifies this potential wavelength
dependence of the degree of polarization (Serkowski 1973; Serkowski et al. 
1975).

Recently, Lazzati et al. (2003) have discussed the potential imprint of cosmic
dust in GRB hosts on the interpretation of polarization data, with particular
emphasis on the afterglow of GRB 021004.  Whereas these authors concentrated
on the interpretation of the time evolution of the percentage polarization,
$P(t)$, here we focus on the wavelength dependence of $P$ at a fixed time. The
latter comes into play if dust dominates the polarimetric properties of the
afterglow light. While the time evolution of $P$ is basically an indicator for
the GRB outflow geometry (Ghisellini \& Lazzati 1999; Sari 1999; Rossi et
al. 2004; Lazzati et al. 2004; for the cannonball model see Dado et al. 2004),
the wavelength dependence of $P$ is a signature for the presence of dust in
the GRB host. Also, whereas a measure of $P(t)$ requires an intense monitoring
of the rapidly fading afterglow over many days, the wavelength dependence of
$P$ can be investigated at any time, in particular when the afterglow is still
in its earliest evolutionary phase, i.e., when the afterglow is brightest.
Rapid multiwavelength polarimetric observations of GRB afterglows could then
allow us to correct for the influence of dust once the time evolution of $P$
is studied.

\section{Theoretical approach}

Our basic assumption is that on average  the dust properties in the GRB host
galaxy resemble those of the dust in our Galaxy. Though  this remains to be
proven, currently it is the best approach for such a study.
In order to describe the wavelength dependence of the percentage
polarization due to dust in the GRB host at a redshift $z$, we adopt the
empirical Serkowski law,  which for a redshifted host in the observer frame 
reads
\begin{equation}
P(\lambda; z)= P_{\rm max} \, \exp(-K \, \mbox{ln}^2\,
((1+z) \ \lambda_{\rm max}^{\rm host}/\lambda))\,.
\label{serkowski}
\end{equation}
We set $K=(1.86 \pm 0.09)\lambda_{\rm max}^{\rm host} (\mu\mbox{\rm m}) -
(0.10 \pm 0.05)$ (Wilking et al. 1982).  In Eq.~(\ref{serkowski})
$\lambda$ stands for the wavelength of observation and $P_{\rm max}$ is the
maximum percentage polarization at the corresponding wavelength $\lambda_{\rm
max}^{\rm host}$ in the host. Following Serkowski et al. (1975), we set
\begin{equation}
\lambda_{\rm max}^{\rm host} (\mu\mbox{m}) = R_V/5.5\,, 
\label{Rv}
\end{equation}
where $R_V$ is the ratio of
total-to-selective extinction. In the case of  spectropolarimetric
observations of a dust-affected afterglow, the observer would simply find a
redshifted (and stretched) Serkowski law (Eq.~\ref{serkowski}). For
broad-band photometric observations in a certain band $X$, however, the
percentage polarization as a function of redshift $z$ is given by
\begin{equation}
P (X; z) = \frac{\int_0^\infty\ d\lambda \ 
            S_X (\lambda)\, F(\lambda)\, P(\lambda; z)}
           {\int_0^\infty \ d\lambda \ S_X (\lambda) \, F(\lambda)}\,.
\label{poleqn}
\end{equation}
Here $\lambda = \lambda_{\rm host}/(1+z)$ is the wavelength of observation,
$F(\lambda)$ is the observed flux density of the afterglow per unit
wavelength, and $S_X(\lambda)$ the filter response function. For $F(\lambda)$
we adopt a synchrotron spectrum modified by extinction in the host:
$F(\lambda) \propto \lambda_{\rm host}^\beta \ \exp(-\tau(\lambda_{\rm
host}))$, where $\beta$ is the spectral index,
$\tau$ is the wavelength dependent optical depth in the host.
When calculating the latter for a given  visual extinction $A_V$(host), we
assumed the extinction curve of the Milky Way (cf. Reichart 2001). For the
filter response functions, $S_X(\lambda)$, we used the data provided by ESO
for VLT/FORS1 and VLT/ISAAC.

\begin{figure}
\centering \includegraphics[width=\columnwidth,angle=0]{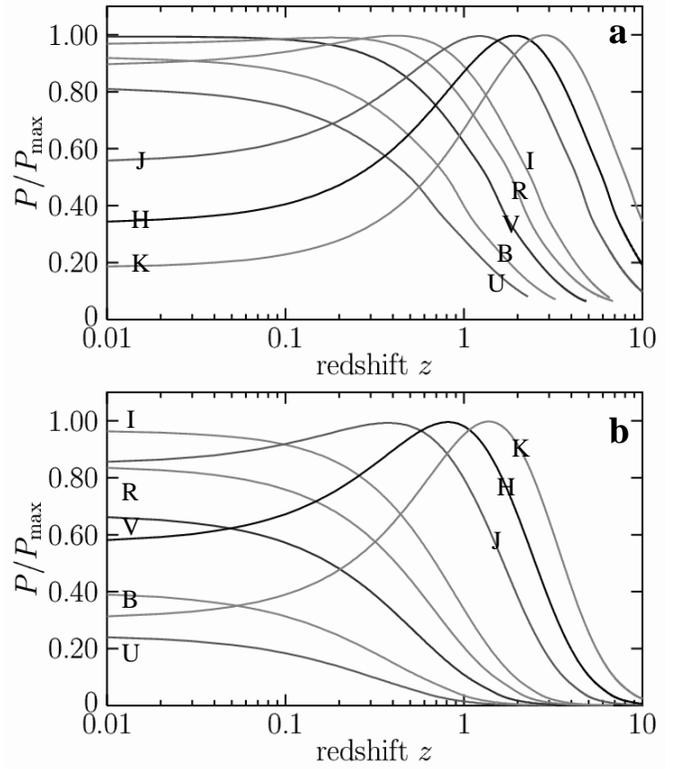}
\caption{The theoretically expected dependence of the degree of linear
polarization of an intrinsically unpolarized GRB afterglow  due to cosmic dust
in the GRB host galaxy as a function of  redshift for the
optical/near-infrared bands (Eq. \ref{poleqn}).  Assumed here is a Milky Way
extinction law with (a) $R_V$=3.1 and (b) $R_V$=5 (Eq.~\ref{Rv}).
The redshifted 2200~\AA \ feature is only marginally evident in these curves.
Nevertheless, we caution that  once the redshift shifts the
photometric bands into the UV domain the predictions become less
secure.}
\label{polofz}
\end{figure}

We further assumed that  the position angle of linear polarization due to
dust, $\theta$, is wavelength independent. This simplification reflects to
some degree our incomplete knowledge about the nature of the star-forming
regions in which the burster might be placed. As long as the alignment of the
scattering grains in the GRB host galaxy is affected by the global magnetic
field acting in the host (cf. Covino et al. 1997),  we might expect however
that those dust grains which are mainly responsible for the NIR polarization
are aligned in the same manner as those which are mainly responsible for the
polarization at shorter wavelengths. This assumption is supported by
observational data  from star-forming regions in our galaxy (e.g., McGregor et
al. 1994; Whittet et al. 1994). However, if the line of sight crosses dense
clumps in interstellar clouds, $\theta$ can become wavelength-dependent
(e.g., Messinger et al. 1997). 

The synchrotron origin of the afterglow light implies that it is linearly
polarized. For our purpose a strict treatment of the propagation of this
radiation is not necessary since its polarization does not depend on
wavelength. The subsequent scattering of the light emerging from the afterglow
by dust grains will alter its polarization state (leading to an increase or
decrease of $P$ depending on both the orientation and
shape of the grains) and leave a spectral imprint which is solely due to the
dust.

Figure~\ref{polofz} displays the percentage polarization we expect to measure
for an intrinsically unpolarized GRB afterglow, if the polarization properties
of the afterglow light are caused by dust in the host  following a Serkowski
law (Eq.~\ref{poleqn}), after the correction for the Galactic dust
contribution. The degree of linear  polarization is plotted as a function of
redshift for the optical/NIR bands in units of the maximum polarization
$P_{\rm max}$ measured by the observer at $\lambda_{\rm max} =
(1+z)\,\lambda_{\rm max}^{\rm host}$.

From Fig.~\ref{polofz} several results become apparent. (1) Polarization in
the NIR becomes higher than in the optical for redshifts $z\gr
0.5...1$. Maximum polarization in the NIR bands occurs  at redshifts $z$=1 to
3.  Since most bursts are at redshifts $z\gr0.5$, NIR polarimetric
observations would be more sensitive for dust in the host in basically all
cases.  (2) Optical versus NIR polarimetry could clearly reveal the presence
of cosmic dust in a GRB host galaxy along the line of sight due to the strong
differences in the (expected) percentage polarization which cannot be
intrinsic to the fireball light. (3) If the intrinsic polarization of GRB
afterglows is usually rather low  at all times, then in some photometric bands
the polarization due to dust in the host can dominate (e.g., in $JHK$),
whereas at the same time in other bands the opposite occurs (e.g., $UBVRI$).
In addition, we found that the functional form of $P(\lambda)/P_{\rm max}$ is
not much sensitive on the assumed visual extinction, $A_V$(host)
(Eq.~\ref{poleqn}), as well as on the spectral index, $\beta$, of the
intrinsic afterglow light.

Naturally, these results rely on the assumption of the Serkowski law,
including the relation for the parameter $K=K(\lambda_{\rm max})$ and the
assumed value for $\lambda_{\rm max}^{\rm host}$ (Eq.~\ref{Rv}).   In
particular, more generalized functions might be used to describe $P(\lambda)$
in the NIR range (e.g., Martin et al. 1999). However, this does not affect the
qualitative trends which are evident in Fig.~\ref{polofz}.

\section{The case GRB 030329}

\begin{figure}
\centering \includegraphics[width=\columnwidth,angle=0]{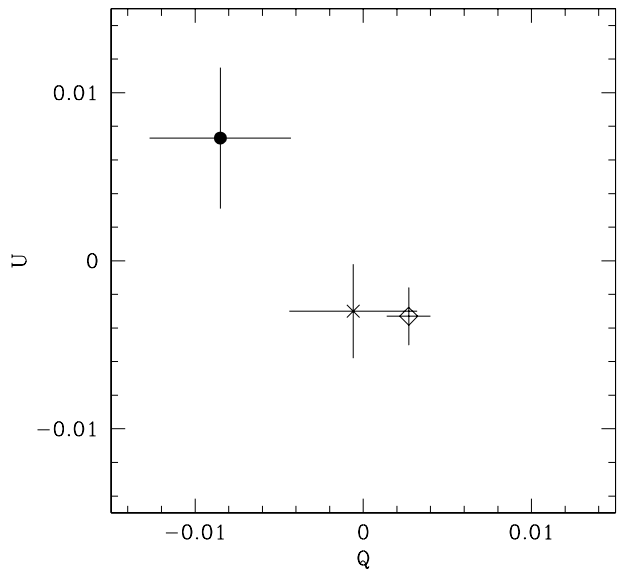}
\caption{Plot showing the values of the Stokes' parameters $U$ and $Q$
including their 1$\sigma$ error bars,  not corrected for ISM polarization, of
the OT  of GRB 030329 (marked with a filled dot)  and of the average of field
stars (cross) in the averaged $R$-band CAFOS  data. The latter agrees well
with the mean value for $U$ and $Q$ of field  stars from the VLT+FORS1
$R$-band campaign on this OT (Greiner et al.  2003a), which is reported as an
open diamond. The OT is separated from the  region occupied by the field
stars: this suggests that it indeed has net  intrinsic polarization, albeit at
low level.}
\label{cafos}
\end{figure}

Optical polarimetry of the Optical Transient (OT) of GRB 030329 was acquired
in the $R$ band with the 2.2-m telescope at Calar Alto (Spain) equipped with
the multi-purpose instrument CAFOS. The 2048$\times$2048 SITe CCD covered a
circular field of diameter 16$'$.  A Wollaston prism with rotatable
half-wavelength, phase-retarder plate allowed splitting the light from each
object in the field into two beams (separated by $\sim20''$ on the CCD frame)
which differed for 90$^\circ$ in their phase angle. By rotating the retarder
plate by 45$^\circ$, one can sample the Stokes' polarization parameters $U$
and $Q$ by using two consecutive images. The quasi-perfect transmission of the
Wollaston prism makes two orientations sufficient for linear polarization
measurements (Lamy \& Hutsem\'ekers 1999).  Using this setup, six pairs of
images, with exposure time of 300 s for each frame, were secured between 22:26
UT of 30 March 2003 and 02:14 UT of 31 March 2003, when the brightness of the
afterglow was $R\approx 16.2$.  The frames were bias-subtracted and
flat-fielded with the standard cleaning procedure, having care of using flat
fields collected using the same rotation angle in the retarder plate as of the
corresponding GRB field image. PSF fitting as implemented in {\sl DAOPHOT II}
(Stetson 1987) within MIDAS was used for the data analysis. The pairs of
images were first analyzed separately, to look for variations in the $U$ and
$Q$ values of OT and field stars during the observations. As none were found,
we coadded together the images with the same retarder plate angle and
considered the average images in order to increase the signal-to-noise  ratio
of the polarization measurement. In order to correct for the instrumental plus
Galactic interstellar medium (ISM)  contribution to the total polarization we
used a number of bright field stars around the OT and as close as possible to
it on the image.  To evaluate the Stokes' $U$ and $Q$ parameters of the
optical transient emission, and its percentage polarization $P$ and position
angle $\theta$, we applied the method described by di Serego Alighieri
(1997). The values (not corrected for the ISM contribution) of these
parameters for the OT and their average  values for the field stars are
reported in Fig.~\ref{cafos}.  The  corrected measurement of $U$ and $Q$ from
our data gives a polarization percentage  $P = 1.1\pm$0.4\% and a polarization
angle $\theta$ = 70$^{\circ}\pm11^\circ$. This value of $P$ is  also corrected
for the polarization bias (Wardle \& Kronberg 1974).

\begin{figure}
\centering \includegraphics[width=\columnwidth,angle=0]{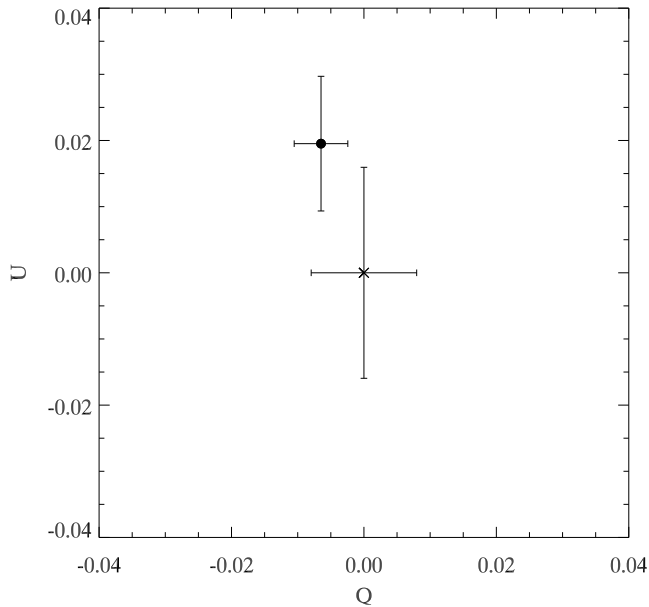}
\caption{The deduced Stokes' parameters $U$ and $Q$ of the afterglow light
in the $K$ band based on the Omega Cass data. The cross
marks the polarization of the field star while the filled dot represents
the afterglow. At the time of the observation the flux from the underlying
host galaxy and from the supernova component (SN 2003dh) can be fully 
neglected.}
\label{ocass}
\end{figure}

\begin{figure}
\centering \includegraphics[width=\columnwidth,angle=0]{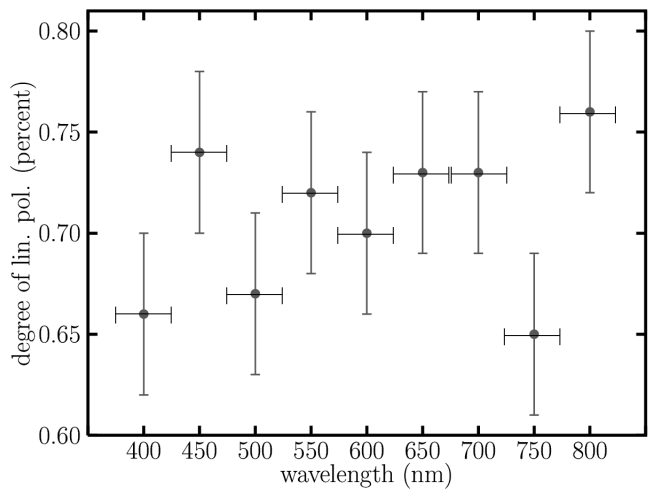}
\caption{Result of the spectropolarimetry of the afterglow of GRB 030329
based on data taken with VLT/FORS1 1.63 days after the burst, nearly
simultaneously with the Calar Alto data. There is no obvious
evidence for a wavelength dependence of $P$.}
\label{vlt}
\end{figure}

$K$-band polarimetry of the afterglow of GRB 030329 was performed with the NIR
camera Omega Cass mounted at the Calar Alto 3.5m telescope on 30.3.2003, UT
22:18 -- 23:18, when the afterglow had a magnitude of $K\approx 14$. Omega
Cass is equipped with a Rockwell $1024\,\times\,1024$ pixel HAWAII array. The
camera was used in the wide-field mode (0.3$''$/pixel), giving a field of view
of $5'\,\times\,5'$. Again, a Wollaston prism was used.  Four polarization
sequences were performed, yielding images at values of 0/90$^\circ$ and
45/135$^\circ$ of the position angle, respectively. There was only one
comparison star in the field, which was assumed to be intrinsically
unpolarized. Thus its net polarization includes polarization due to both the
ISM and the instrument, and defines the zero point for the calculation of the
afterglow polarization. From Fig.~\ref{ocass} it is obvious that the error of
the comparison star which is the brightest star in the field of view of Omega
Cass dominates the error budget since it was about 0.3 mag fainter than the
afterglow. The final result is $P=2.1\pm 1.2\%, \theta = 54.1^\circ\pm
10.4^\circ$ (Fig.~\ref{ocass}), after correction for the polarization bias.
Even though our measurement error allows for a $P=0$ (within 2$\sigma$), we
note that the polarization angle we deduce for the $K$ band is basically
identical to those found for the $R$ band. This makes us confident
that our measurement relies on a real detection and is not an upper limit.

In principle, the amount of the percentage polarization measured in $R$ and
$K$ could be used to constrain the extinction in the host along the line of
sight using the empirical  relation $P_{\rm max} \kr 9$ $E$(B-V) found for
Galactic dust (Schmidt-Kaler 1958; Serkowski et al. 1975).  However, the less
known ratio of total-to-selective extinction of the dust in the host, the
scatter of the $P_{\rm max}$ vs. $E(B-V)$ relation for small $E(B-V)$
(cf. Serkowski et al. 1975), as well as the observed spread in the  empirical
parameter $K=K(\lambda_{\rm max})$ (Eq.~\ref{serkowski}; cf. Wilking et
al. 1982; Whittet et al. 1992) naturally limits the validity   of such an
approach.

A qualitative constraint on the impact of dust  in the GRB host  galaxy on the
fireball light can be set by comparing the predicted ratios of the percentage
polarization in the different photometric bands (Fig.~\ref{polofz})  with the
observed ones. For $R_V=3.1$ ($R_V=5$) at the redshift of GRB 030329
($z$=0.1685; Greiner et al. 2003b), the predicted ratio of the percentage
polarization between the $R$- and the $K$ band is approximately 4.0 (1.6) if
dust dominates, and 1.0 if there is no (aligned) dust. Since the former is not
evident in our data (we measure $P_R/P_K = 0.5 \pm 0.3$), we conclude that at
the time of our observations, $\sim$1.5 days after the burst, the intrinsic
polarization  of the afterglow dominates over any polarization  possibly
introduced by dust in the GRB host galaxy.  This conclusion is strengthened by
the results of the VLT polarimetry of the afterglow of GRB 030329 (Greiner et
al. 2003a). Spectropolarimetric VLT data obtained 1.63 days after the burst,
thus nearly simultaneously with the Calar Alto data,  do not show evidence for
a smooth wavelength dependence of $P$ in the region between 400 and 800~nm
(Fig.~\ref{vlt}). The same holds for the spectropolarimetric VLT data taken
0.64, 2.56, and 2.64 days after the burst (Greiner et al., in preparation).

\section{Concluding remarks}

Encouraged by our successful simultaneous optical and NIR polarimetric
observations of the afterglow of GRB 030329, we have explored here in which
manner simultaneous multicolor polarimetric observations can reveal an imprint
from cosmic dust in GRB host galaxies on the afterglow light.  We have argued
that dust in GRB hosts can strongly affect the polarization properties of an
afterglow, depending on its redshift and on the photometric band, in which the
observations are performed. In particular, the wavelength dependence of
polarization due to dust, which is in strong contrast to the intrinsic
wavelength independence of the fireball light, could be used as an indicator
for the presence of dust. 

With the upcoming $Swift$ satellite mission GRB afterglows are expected to be
localized within seconds after burst trigger, in principle allowing for first
polarimetric observations within the first hour after a burst.  Since the
amount of  maximum polarization, $P_{\rm max}$, can rapidly increase with
increasing visual extinction within the host galaxy (Schmidt-Kaler 1958;
Serkowski et al. 1975), in principle large degrees of linear polarization
could be detectable for the most  intrinsically extinguished
afterglows. Conversely, if a wavelength dependent polarization is never found
in an afterglow then either the  host dust properties must be substantially
different from those of the Galactic ISM, or there is no dust at all, or
the magnetic field in the host galaxy is weak and/or not well ordered.

The position angle of the polarization defines the orientation of the
predominant magnetic field in the host galaxy along the line of sight.  This
predominant field can either be the large scale magnetic field acting in the
GRB host (in particular, if the host is seen edge-on), or the field
characterizing the star-forming region in which the burster is placed.
The detection of interstellar dust extinction in the host
without a substantial polarization signal could hint to a strongly tangled
magnetic field structure in the host galaxy, or even indicate the absence of a
magnetic field  large enough ($\sim 10^{-8}$~G) to align the dust grains in
the host galaxy. Both effects would indicate very strong evolutionary effects
in the galactic magnetic field structure as practically all well studied
nearby galaxies are known to have ordered magnetic field on large ($\sim$~kpc)
scales (Kronberg 1994; Scarrott 1996; Wielebinski \& Krause 1993).
In the Milky Way the dust grain alignment along the arbitrary line of
sights due to the Galactic magnetic field produces a polarization per
magnitude of visual extinction in the range $P/A_{\rm V}\approx 0.6-3~\%\,{\rm
mag}^{-1}$ (Serkowski et al. 1995). Thus a polarization below this value might
indicate special conditions in a GRB host galaxy, including the star-forming
region in which the burster is placed.

\begin{acknowledgements}

S.K. acknowledges support from the Deutsche Akademische Austauschdienst (DAAD)
under grants No. D/0237747 and D/0103745; N.M. and E.P. acknowledge support
under CRUI  Vigoni program 31-2002. D.H.H. acknowledges support under NSF
grant INT-0128882. This work is partly based on observations at the
German-Spanish Astronomical Centre, Calar Alto, operated by the
Max-Planck-Institute for Astronomy, Heidelberg, jointly with the Spanish
National Commission for Astronomy. We thank the staff at Calar Alto for their
assistence, in particular A. Aguirre, M. Alises, S. Pedraz, and U. Thiele.  We
thank Elena Pian for a careful reading of the manuscript. We thank the
referee for a rapid response and helpful remarks.

\end{acknowledgements}


\end{document}